\documentstyle[12pt]{article}
\begin{document}

\begin{titlepage}

\title{\bf PARK--TARTER MATRIX FOR A DYON--DYON SYSTEM}

\author{
L. G. Mardoyan${}^1$,
A. N. Sissakian,
V. M. Ter--Antonyan${}^2$ }
\date{\nonumber}

\maketitle
\footnotetext[1]{E-mail: mardoyan@thsun1.jinr.dubna.su}
\footnotetext[2]{E-mail: terant@thsun1.jinr.dubna.su}

\begin{abstract}
The problem of separation of variables in a dyon--dyon system is discussed.
A linear transformation is obtained between fundamental bases of this
system. Comparison of the dyon--dyon system with a 4D isotropic oscillator
is carried out.
\end{abstract}

\thispagestyle{empty}
\end{titlepage}

\section{Introduction}
In this paper, we have calculated the matrix between the spherical and
parabolic bases of a dyon--dyon system [1] belonging to the same energy
level. This matrix is a generalization of the Park--Tarter matrix known
from the theory of hydrogen atom [2, 3] to the case when the Coulomb
center carries not only the electric but also magnetic charge. Like the
Park--Tarter matrix, our matrix is expressed through the Clebsch--Gordan
coefficients $C_{a\alpha;b\beta}^{c\gamma}$, however, in our case
$a \ne b$, in contrast to the case of a
hydrogen atom. We have also traced the connection of the dyon--dyon problem
with that of a 4-dimensional isotropic oscillator. As is known [4],
these problems are related to each other by the Kustaanheimo--Stiefel
transformation [5] supplemented with the 4th (angular) coordinate. We
have shown that the coefficients $C_{a\alpha;b\beta}^{c\gamma}$ coincide
with the ones [6] of the expansion of the double polar basis over the
Euler basis of a 4-dimensional isotropic oscillator.

\section{Dyon--Dyon System}
A dyon--dyon system in the space ${\rm I \!R}^3$ is described by the equation
\begin{eqnarray}
\left[\left(\frac{\partial}{\partial x_j}-
\frac{ie}{\hbar c}A_j\right)^2 - \frac{s^2}{r^2}\right]\psi
+ \frac{2M_0}{\hbar^2}\left(\epsilon^s + \frac{e^2}{r}\right)\psi = 0
\label{eq:1}
\end{eqnarray}
where
\begin{eqnarray*}
A_j = \frac{gx_3}{r(r^2-x_3^2)}(-x_2,x_1,0)
\end{eqnarray*}
and $s=eg/\hbar c = 0, \pm 1/2, \pm 1,...$.
Each value of $s$ describes its particular dyon--dyon system.
At $s=0$, eq.(1) is reduced to the Schr\"odinger equation for a
hydrogen atom. When $s \ne 0$, equation (1) preserves O(4)-symmetry
and therefore variables in it are separated into spherical, parabolic,
and prolate spheroidal coordinates [1].

The system (1) possesses a singularity on the axis $x_3$. It is also
possible to consider systems with singularities either on the semiaxis
$x_3 > 0$ or on $x_3 < 0$,
i.e. they are described by the vector potentials
\begin{eqnarray*}
A_j^{(\pm)} = \frac{g}{r(r\mp x_3)}(\mp x_2,\pm x_1,0)
\end{eqnarray*}
and are connected with the system (1) by the gauge transformations
\begin{eqnarray*}
A_j^{(\pm)} = A_j  + \frac{\partial f^{(\pm)}}{\partial x_j},
\,\,\,\,\,\,\,\,\,\,\,\,
\psi^{(\pm)}(\vec x) = \psi(\vec x)
\exp\left({\frac{ie}{\hbar c}f^{(\pm)}}\right)
\end{eqnarray*}
with the gauge function $f^{(\pm)} = \pm 2g \arctan{x_2/x_1}$.

The variables in eq. (1) are separated in spherical and parabolic coordinates.

In the spherical coordinates
\begin{eqnarray}
x_1 = r\sin \theta \cos \varphi,\,\,\,\,\,\,\,\,\,\,\,\,
x_2 = r\sin \theta \sin \varphi,\,\,\,\,\,\,\,\,\,\,\,\,
x_3 = r\cos \theta
\label{eq:2}
\end{eqnarray}
the wave function of the dyon--dyon system is of the form [7]
\begin{eqnarray*}
\psi_{nkm}^{(s)}(r,\theta,\varphi) = R_{nkm}^{(s)}(r)
Z_{km}^{(s)}(\theta)
\frac{e^{im\varphi}}{\sqrt{2\pi}}
\end{eqnarray*}
where the functions $Z_{km}^{(s)}(\theta)$ and $R_{nkm}^{(s)}(r)$
normalized by the condition
\begin{eqnarray*}
\int \limits_{0}^{\pi} \sin \theta
Z_{k'm}^{(s)}(\theta)Z_{km}^{(s)}(\theta)d\theta = \delta_{k'k}
,\,\,\,\,\,\,\,\,\,\,
\int \limits_{0}^{\infty} r^2 \left[R_{nkm}^{(s)}(r)\right]^2dr = 1
\end{eqnarray*}
are given by the formulae
\begin{eqnarray*}
Z_{km}^{(s)}(\theta) = N_{km}^{(s)}(1-\cos \theta)^{\frac{|m-s|}{2}}
(1+\cos \theta)^{\frac{|m+s|}{2}}
P_k^{(|m-s|,|m+s|)}(\cos \theta)
\end{eqnarray*}
\begin{eqnarray*}
R_{nkm}^{(s)}(r) = C_{nkm}^{(s)}
\exp{\left(-\frac{r}{r_0n}\right)}
\left(\frac{2r}{r_0n}\right)^{k+\frac{|m-s|+|m+s|}{2}} \nonumber \\ [3mm]
F\left(-n+k+\frac{|m-s|+|m+s|}{2}+1; 2k+|m-s|+|m+s|+2;
\frac{2r}{r_0n}\right)
\end{eqnarray*}
Here $P_n^{(\alpha,\beta)}(x)$ are Jacobi polynomials; $r_0=\hbar^2/M_0e^2$
is the Bohr radius. The normalization
constants $N_{km}^{(s)}$ and $C_{nkm}^{(s)}$ equal
\begin{eqnarray*}
N_{km}^{(s)} = \left[\frac{(2k+|m-s|+|m+s|+1)k!(k+|m-s|+|m+s|)!}
{2^{|m-s|+|m+s|+1}\Gamma(k+|m-s|+1)\Gamma(k+|m+s|+1)}\right]^{1/2}
\end{eqnarray*}
\begin{eqnarray*}
C_{nkm}^{(s)} = \frac{2}{n^2r_0^{3/2}}\frac{1}{(2k+|m-s|+|m+s|+1)!}
\sqrt{\frac{\left(n+k+\frac{|m-s|+|m+s|}{2}\right)!}
{\left(n-k-\frac{|m-s|+|m+s|}{2}-1\right)!}}
\end{eqnarray*}
Quantum numbers run over the values $n=1,3/2,2,..., k=0,1,...k_{max}$, where
\begin{eqnarray*}
k_{max} = n-\frac{|m-s|+|m+s|}{2} - 1
\end{eqnarray*}
The energy spectrum of the system is of the form
\begin{eqnarray}
\epsilon_n^s = -\frac{M_0e^4}{2\hbar^2n^2}
\label{eq:3}
\end{eqnarray}

In the parabolic coordinates
\begin{eqnarray}
x_1 = \sqrt{\xi \eta}\cos \varphi,\,\,\,\,\,\,\,
x_2 = \sqrt{\xi \eta}\sin \varphi,\,\,\,\,\,\,\,
x_3 = \frac{1}{2}(\xi - \eta)
\label{eq:4}
\end{eqnarray}
upon the substitution
\begin{eqnarray*}
\psi(\xi,\eta,\varphi) = f_1(\xi)f_2(\eta)
\frac{e^{im\varphi}}{\sqrt{2\pi}}
\end{eqnarray*}
the variables in (1) are separated, which results in the system of equations
\begin{eqnarray*}
\frac{d}{d\xi}\left(\xi\frac{df_1}{d\xi}\right) +
\left[\frac{M_0\epsilon^s}{2\hbar^2}\xi - \frac{(m+s)^2}{4\xi}
+ \beta_1\right]f_1 = 0 \\ [3mm]
\frac{d}{d\eta}\left(\eta\frac{df_2}{d\eta}\right) +
\left[\frac{M_0\epsilon^s}{2\hbar^2}\eta - \frac{(m-s)^2}{4\eta}
+ \beta_2\right]f_2 = 0
\end{eqnarray*}
where
\begin{equation}
\beta_1 + \beta_2 = \frac{M_0e^2}{\hbar^2}
\label{eq:5}
\end{equation}
At $s = 0$, these equations coincide with the equations for a hydrogen
atom in the parabolic coordinates [8], and consequently,
\begin{eqnarray*}
\psi_{n_1n_2m}^{(s)}(\xi,\eta,\varphi) = \frac{\sqrt{2}}{n^2r_0^{3/2}}
f_{n_1,m+s}(\xi)f_{n_2,m-s}(\eta)\frac{e^{im\varphi}}{\sqrt{2\pi}}
\end{eqnarray*}
where
\begin{eqnarray*}
f_{pq}(x) = \frac{1}{\Gamma(|q|+1)}
\sqrt{\frac{\Gamma(p+|q|+1)}{p!}}
\exp{\left(-\frac{x}{2r_0n}\right)}
\left(\frac{x}{r_0n}\right)^{\frac{|q|}{2}}
F\left(-p; |q|+1; \frac{x}{r_0n}\right)
\end{eqnarray*}
Here $n_1$ and $n_2$ are non-negative integers
\begin{eqnarray*}
n_1 = -\frac{|m+s|+1}{2} + \frac{\hbar}{\sqrt{-2M_0\epsilon^s}}\beta_1
,\,\,\,\,\,\,\,\,\,\,
n_2 = -\frac{|m-s|+1}{2} + \frac{\hbar}{\sqrt{-2M_0\epsilon^s}}\beta_2
\end{eqnarray*}
from which and (3), (5) it follows that the parabolic quantum numbers
$n_1, n_2, m$ and $s$ are connected with the principal quantum number
$n$ as follows:
\begin{eqnarray}
n = n_1 +n_2 + \frac{|m-s|+|m+s|}{2} + 1
\label{eq:6}
\end{eqnarray}
\section{Park--Tarter Generalized Matrix}
We write the searched expansion in the form
\begin{eqnarray}
\psi_{n_1n_2m}^{(s)}(\xi,\eta,\varphi) = \sum_{k=0}^{k_{max}}
T_{n_1n_2km}^{(s)}
\psi_{nkm}^{(s)}(r,\theta,\varphi)
\label{eq:7}
\end{eqnarray}
Our purpose is to calculate the coefficients $T_{n_1n_2km}^{(s)}$, i.e.
the Park--Tarter generalized matrix. The usual Park--Tarter matrix is the
matrix $T_{n_1n_2km}^{(s)}$ at $s = 0$.

We substitute
\begin{eqnarray*}
\xi = r(1 + \cos\theta),\,\,\,\,\,\,\,\,
\eta = r(1 - \cos\theta),\,\,\,\,\,\,\,\,
\end{eqnarray*}
into the left-hand side of expansion (7), let $r$ tend to infinity, take
the formula
\begin{eqnarray*}
F(-n;c;x) \sim (-1)^n\frac{\Gamma(c)}{\Gamma(c+n)}x^n,
\,\,\,\,\,\,\,\,\,\,\,\,\,\, (x \to \infty)
\end{eqnarray*}
and the orthogonality condition for the function $Z_{km}^{(s)}$ into account. All this
leads to the formula
\begin{eqnarray*}
T_{n_1n_2km}^{(s)} = (-1)^k B_{n_1n_2km}^{(s)}I_{n_1n_2km}^{(s)}
\end{eqnarray*}
where
\begin{eqnarray*}
B_{n_1n_2km}^{(s)} = \sqrt{\frac{(2k+|m-s|+|m+s|+1)k!(k+|m-s|+|m+s|)!}
{2^{2n+|m-s|+|m+s|}\Gamma(k+|m-s|+1)\Gamma(k+|m+s|+1)}}
\\ [3mm]
\left[\frac{\left(n-k-\frac{|m-s|+|m+s|}{2}-1\right)!
\left(n+k+\frac{|m-s|+|m+s|}{2}\right)!}
{(n_1)!(n_2)!
\Gamma(n_1+|m+s|+1)\Gamma(n_2+|m-s|+1)}\right]^{1/2}
\end{eqnarray*}
and the second factor is equal to the integral
\begin{eqnarray*}
I_{n_1n_2km}^{(s)} = \int\limits_{-1}^{1}
(1-x)^{n_2+|m-s|}(1+x)^{n_1+|m+s|}
P_k^{(|m-s|,|m+s|)}(x)dx
\end{eqnarray*}
Then taking advantage of the Rodrigues formula [9]
\begin{eqnarray*}
P_{n}^{(\alpha,\beta)}(x) = \frac{(-1)^n}{2^nn!}(1-x)^{-\alpha}
(1+x)^{-\beta} \frac{d^n}{dx^n}
\left[(1-x)^{\alpha + n}(1+x)^{\beta + n}\right]
\end{eqnarray*}
and the integral representation for the Clebsch--Gordan coefficients [10]
\begin{eqnarray*}
C_{a\alpha;b\beta}^{c\gamma} = \delta_{\alpha + \beta = \gamma}
\left[\frac{(2c+1)(J+1)!
(J-2c)!(c+ \gamma)!}{(J-2a)!(J-2b)!(a- \alpha)!(a+ \alpha)!
(b- \beta)!(b+ \beta)!(c- \gamma)!}\right]^{1/2} \\ [3mm]
\frac{(-1)^{a-c+\beta}}{2^{J+1}}
\int\limits_{-1}^{1}(1-x)^{a-\alpha}(1+x)^{b-\beta}
\frac{d^{c-\gamma}}{dx^{c-\gamma}}
\left[(1-x)^{J-2a}(1+x)^{J-2b}\right]dx
\end{eqnarray*}
($J=a+b+c$), we obtain
\begin{eqnarray}
T_{n_1n_2ms}^{nk} = (-1)^{n_2+k}
C_{a\alpha;b\beta}^{c\gamma}
\label{eq:8}
\end{eqnarray}
where
\begin{eqnarray*}
a=\frac{n_1+n_2+|m+s|}{2},\,\,\,\,\,\,\,
b=\frac{n_1+n_2+|m-s|}{2},\,\,\,\,\,\,\,
c=k+\frac{|m-s|+|m+s|}{2}  \\[3mm]
\alpha=\frac{n_1-n_2+|m+s|}{2},\,\,\,\,\,\,\,
\beta=\frac{n_2-n_1+|m-s|}{2},\,\,\,\,\,\,\,
\gamma=\frac{|m-s|+|m+s|}{2}
\end{eqnarray*}
At $s = 0$ formula (8) turns into the Park--Tarter formula, as would be
expected.
\section{Dyon--Dyon System and 4D Oscillator}
Let us demonstrate that if in eq. (1) we make the changes
\begin{eqnarray}
s \to -i\frac{\partial}{\partial \gamma},\,\,\,\,\,\,\,\,\,\,\,\,
\psi(\vec x) \to \psi(\vec x,\gamma) = \psi(\vec x)
\frac{e^{is\gamma}}{\sqrt{4\pi}}
\label{eq:9}
\end{eqnarray}
($\gamma \in [0, 4\pi)$), it will transform into the Schroedinger equation
for a 4D isotropic oscillator.

Equation (1) in the spherical coordinates is of the form
\begin{eqnarray}
\frac{1}{r^2}\frac{\partial}{\partial r}\left(
r^2\frac{\partial \psi}{\partial r}\right) + \frac{1}{r^2}\left[
\frac{1}{\sin \theta}\frac{\partial}{\partial \theta}\left(
\sin \theta \frac{\partial \psi}{\partial \theta}\right) +
\frac{1}{\sin^2\theta}\frac{\partial^2\psi}
{\partial \varphi^2}\right] - \nonumber \\ [3mm]
\frac{2is\cos \theta}{r^2\sin^2\theta}\frac{\partial \psi}{\partial \varphi} -
\frac{s^2}{r^2\sin^2\theta}\psi +
\frac{2M_0}{\hbar^2}\left(\epsilon^s + \frac{e^2}{r}\right)\psi = 0
\label{eq:10}
\end{eqnarray}
From (9) and (10) we have
\begin{eqnarray}
\left[\frac{1}{r^2}\frac{\partial}{\partial r}\left(r^2
\frac{\partial}{\partial r}\right) - \frac{{\hat J}^2}{r^2}\right]
\psi + \frac{2M_0}{\hbar^2}\left(\epsilon^s + \frac{e^2}{r}\right)
\psi = 0
\label{eq:11}
\end{eqnarray}
where
\begin{eqnarray*}
{\hat J}^2 = -\left[\frac{1}{\sin \beta}\frac{\partial}{\partial \beta}
\left(\sin \beta\frac{\partial}{\partial \beta}\right) +
\frac{1}{\sin^2\beta}\left(\frac{\partial^2}{\partial \alpha^2} -
2\cos\beta\frac{\partial^2}{\partial \alpha \partial \gamma}+
\frac{\partial^2}{\partial \gamma^2}\right)\right]
\end{eqnarray*}
Here we change the notation: $\beta = \theta$ and $\alpha = \varphi$.
If we now pass from the coordinates $r, \alpha, \beta, \gamma$ to the
coordinates
\begin{eqnarray}
u_0 + iu_1 = u\cos{\frac{\beta}{2}}
e^{-i\frac{\alpha + \gamma}{2}},\,\,\,\,\,
u_2 + iu_3 = u\sin{\frac{\beta}{2}}
e^{i\frac{\alpha - \gamma}{2}}
\label{eq:12}
\end{eqnarray}
with $u^2 = r$, take into account that
\begin{eqnarray*}
\frac{{\partial}^2}{\partial u_{\mu}^2} =
\frac{1}{u^3}\frac{\partial}{\partial u}
\left(u^3\frac{\partial}{\partial u}\right) - \frac{4}{u^2}{\hat J}^2
\end{eqnarray*}
and introduce the notation
\begin{eqnarray*}
E = 4e^2,\,\,\,\,\,\,\,\,\,\,\,\,\,\,\,\,\,\,\,\,\,\,\,\,
\epsilon^s = -\frac{M_0\omega^2}{8}
\end{eqnarray*}
then equation (11) will turn into the Schr\"odinger equation for a 4D
isotropic oscillator
\begin{eqnarray*}
\left[\frac{{\partial}^2}{\partial u_{\mu}^2} +
\frac{2M_0}{\hbar}\left(E - \frac{M_0\omega^2 u^2}{2}\right)\right]
\psi(\vec u) = 0
\end{eqnarray*}
whose energy spectrum is given by the formula
\begin{eqnarray}
E_N = \hbar \omega (N+2)
\label{eq:13}
\end{eqnarray}

Introducing the double polar coordinates
\begin{eqnarray}
u_0 + iu_1 = \rho_1 e^{-i\varphi_1},\,\,\,\,\,
u_2 + iu_3 = \rho_2 e^{i\varphi_2}
\label{eq:14}
\end{eqnarray}
from formulae (2), (4), (12), and (14) we get the relations
\begin{eqnarray*}
\xi = 2\rho_1^2,\,\,\,\,\,\,\,\,\,\,\,\,\,\,\ \eta = 2\rho_2^2,
\,\,\,\,\,\,\,\,\,\,\,\,\,\,\,\varphi=\varphi_1+\varphi_2,
\,\,\,\,\,\,\,\,\,\,\,\,\,\,\,\gamma=\varphi_1-\varphi_2
\end{eqnarray*}
which lead to the formulae
\begin{eqnarray*}
\psi_{NJM_1M_2}(u,\alpha,\beta,\gamma) = 4n\sqrt{\frac{2}{\lambda}}
\delta_{n,\frac{N}{2}+1}
\delta_{k,J-\frac{|M_1-M_2|+|M_1+M_2|}{2}}
\delta_{m,M_1}\delta_{s,M_2}
\psi_{nkms}(r,\theta,\varphi,\gamma)
\end{eqnarray*}
\begin{eqnarray*}
\psi_{N_1N_2m_1m_2}(\rho_1,\rho_2,\varphi_1,\varphi_2) =
4n\sqrt{\frac{2}{\lambda}}
\delta_{n_1,N_1} \delta_{n_2,N_2}\delta_{m,\frac{m_1+m_2}{2}}
\delta_{s,\frac{m_1-m_2}{2}}
\psi_{n_1n_2ms}(\xi,\eta,\varphi,\gamma)
\end{eqnarray*}
generalizing the earlier results [6, 11].

Now we are able to write the expansion [6]
\begin{eqnarray}
\psi_{N_1N_2m_1m_2}(\rho_1,\rho_2,\varphi_1,\varphi_2) =
\sum_{J=J_{min}}^{N/2}W_{N_1N_2m_1m_2}^{NJM_1M_2}
\psi_{NJM_1M_2}(u,\alpha,\beta,\gamma)
\label{eq:15}
\end{eqnarray}
where
\begin{eqnarray}
W_{N_1N_2m_1m_2}^{NJM_1M_2} = e^{i\pi \Phi}
C_{a_0,\alpha_0;b_0,\beta_0}^{c_0,\gamma_0}
\label{eq:16}
\end{eqnarray}
\begin{eqnarray*}
a_0 = \frac{N+|m_1|-|m_2|}{4},\,\,\,\,\,\,\,\,\,
b_0 = \frac{N-|m_1|+|m_2|}{4},\,\,\,\,\,\,\,\,\,
c_0 = J \nonumber \\ [3mm]
\alpha_0 = \frac{N+|m_1|-|m_2|}{4}-N_2,\,\,\,
\beta_0 = \frac{N-|m_1|+|m_2|}{4} - N_1,\,\,\,
\gamma_0 = \frac{|m_1|+|m_2|}{2}
\end{eqnarray*}
The lower limit of summation in (15) and quantity $\Phi$ are given
by the expressions
\begin{eqnarray*}
J_{min}=\frac{1}{2}\left(|M_1-M_2|+|M_1+M_2|\right)
\end{eqnarray*}
\begin{eqnarray*}
\Phi = N_2 + J - \frac{|m_1|+|m_2|}{2} - \frac{m_2+|m_2|}{2}
\end{eqnarray*}

We conclude with the following two comments: \\
(a) Using formulae (2) and (12) and considering that
$r=u^2, \theta=\beta, \varphi=\alpha$, one can easily show that
\begin{eqnarray*}
x_1 &=& 2(u_0u_2 + u_1u_3) \\ [3mm]
x_2 &=& 2(u_0u_3 - u_1u_2) \\ [3mm]
x_3 &=& u_0^2 + u_1^2 - u_2^2 - u_3^2 \\ [3mm]
\gamma &=& \frac{i}{2}\ln{\frac{(u_0+iu_1)(u_2+iu_3)}
{(u_0-iu_1)(u_2-iu_3)}}
\end{eqnarray*}
The first three lines are the transformation
${\rm I \!R}^4 \to {\rm I \!R}^3$ suggested by Kustaanheimo
and Stiefel for the regularization of equations of celestial mechanics [5].
Later, this transformation found other applications, as well [12, 13].
This transformation supplemented with the coordinate $\gamma$ was used for
the ''synthesis'' of the dyon--dyon system from the 4D isotropic oscillator
[4]. \\
(b) It is known [6] that diagonal ($m_1 = m_2$) elements of the matrix
$W_{N_1N_2m_1m_2}^{NJM_1M_2}$ with $N$
even coincide with the Park--Tarter matrix. From formula (16) it follows that
the remaining elements of the matrix $W_{N_1N_2m_1m_2}^{NJM_1M_2}$ have
also a physical meaning: these are elements of the generalized
Park--Tarter matrix for the dyon--dyon system.

\section{Degeneracy of the Energy Levels}
Let us discuss the problem of multiplicity of degeneration of the energy
levels (3) and (13). From formula (6) it follows that at fixed $n, m$ and
$s$ the energy levels are degenerate with the multiplicity
\begin{eqnarray*}
g_{nm}^s = n - \frac{|m-s|+|m+s|}{2}
\end{eqnarray*}
For $s \geq 0$ the multiplicity of degeneration of levels (3) at fixed
$s$ and $n$ is
\begin{eqnarray*}
g_{n}^s = \sum_{|m|\geq [s]}g_{nm}^s +
\sum_{|m|\leq [s]-1}g_{nm}^s
\end{eqnarray*}
where the upper limit of summation is determined from the condition
$g_{nm}^s \geq0$,
\begin{eqnarray*}
|m-s|+|m+s| \leq 2n - 2
\end{eqnarray*}
Therefore,
\begin{eqnarray}
g_{n}^s = \sum_{m=-[s]+1}^{[s]-1}(n - s) +
2\sum_{m=[s]}^{[n]-1}(n - m) = (n-s)(n+s)
\label{eq:17}
\end{eqnarray}
The same result follows from analogous computations also when
$s < 0$.

The quantum numbers $s$ and $n$ in formula (17) assume simultaneously either
integer or half-integer values, and thus, we have
\begin{eqnarray*}
g_n = \sum_{s=-n+1}^{n-1} g_{n}^s = \frac{1}{3}n(2n-1)(2n+1)
\end{eqnarray*}
where $g_n$ stands for the multiplicity of degeneration of the energy levels
(13) of the 4d oscillator.
Since $N = 2n -2$, we arrive at the known result
\begin{eqnarray*}
g_N = \frac{1}{6}(N+1)(N+2)(N+3)
\end{eqnarray*}

\begin{center}
{\bf Acknowledgments}
\end{center}
We are sincerely grateful to G.S.Pogosyan for useful
discussions.

\end{document}